\documentclass[12pt]{amsart}
\usepackage[mathscr]{eucal}
\usepackage{amscd}
\usepackage{amsfonts}
\usepackage{amsmath, amsthm, amssymb}
\usepackage{latexsym}
\usepackage[all]{xy}
\usepackage[dvips]{graphics}
\usepackage{anyfontsize}                                                \usepackage{amsmath}                                                                          \usepackage{booktabs}                                                                   \usepackage{titlesec}

\titleformat{\section}
  {\normalfont\Large\bfseries}
  {\thesection}{1em}{}                      
\usepackage{hyperref}
\hypersetup{
colorlinks=true,
linkcolor=blue,
filecolor=blue,
citecolor = black,
urlcolor=blue}
\usepackage{float}

\theoremstyle{plain}
\newtheorem{thm}{Theorem}

\newtheorem{definition}{Definition}[section]
\theoremstyle{definition}

\newtheorem{ex}[thm]{Example}

\theoremstyle{remark}

\newtheorem{lemma}{Lemma}[section]
\newtheorem{proposition}{Proposition}[section]
\newtheorem{theorem}{Theorem}[section]
\newtheorem{remark}{Remark}

\makeatletter
\def\thmhead@plain#1#2#3{%
  \thmname{\bfseries #1}\thmnumber{ #2}\thmnote{ (#3)}}
\let\thmhead\thmhead@plain
\makeatother

\theoremstyle{remark}

\numberwithin{equation}{section}

\usepackage{enumitem}

\title{\bfseries\large Various Kinds of Defining Sets over
$\mathbb{Z}_2^\alpha\mathbb{Z}_4^\beta$
and their $Lee$-Weight distributions}

\author{
S. Das$^{1}$ and M.~K.~Patel$^{2}$\\[1ex]
$^{1,2}$Department of Mathematics,\\
National Institute of Technology Nagaland,\\
Chumoukedima--797103, Nagaland, India
}

\date{}
\begin{document}
 \thispagestyle{empty}

\vspace*{-1.5cm}

\maketitle

\textbf{Abstract}:  Recently, a particular type of defining set is being used to get $Lee$-weight distribution over a particular type of ring. However, little work has been done on mixed alphabets. Therefore, in Section 3, we investigate the $Lee$-weight distributions associated with various kinds of defining sets, labelled A to J, for code constructions over the mixed-alphabet ring $R=Z^{\alpha}_2 Z^{\beta}_4$, where $\alpha$ and $\beta$ are positive finite integers. Using Parseval's identity, we determine the cardinalities of the defining sets A to J. We further establish several new results concerning 2-$Lee$-weight and 3-$Lee$-weight distributions.  \\ 

\textbf{Mathematics Subject Classification}: 81P70, 94B05, 94B15.

\textbf{Keywords:}  Additive codes, Gray map, N-weight codes.
\\\\

\section{Introduction}
Research on $Z_2Z_4$-additive codes starts in late 1990s. In 2001, Borges et al. began systematic study of additive codes over mixed alphabets $Z_2 Z_4$ (see \cite{1}) and later in 2007, they developed its different algebraic properties including duality and their polynomial representations and extended their work using poset (see \cite{5}). Zonbing et al. \cite{2} introduced new family of quaternary codes $C^\delta_{k_1,k_2}$ for $\delta=1,2,3$ which gives two lee weight codes. In \cite{3}, Wu et al. explored four $Lee$ weight quaternary code and established two infinite families of new kinds. Bilal et al.\cite{4} introduced codes of the type $(\alpha,\beta;\gamma,\delta;\kappa)$ which is a family of $Z_2Z_4$-additive code and it is useful to further study additive codes over mixed alphabets of $Z_2$ and $Z_4$. Again duality and weight enumerator for mixed alphabet were studied by Borges et al. \cite{6} and establishes characterization results for self-dual mixed-alphabet codes. In \cite{7}, Geng et al. also employed a Gray map over mixed alphabets and investigated weight enumerators, thereby extending several existing results and later their works were extended by Liu et al \cite{10} for two and three Lee weight codes over non chain rings and they explicitly determined the Lee weight distributions of the constructed codes using Gauss sums over finite fields. Kumar et al. \cite{8}  got few weight codes using defining set $
D=\left\{a+ub\in R_m^*:
\operatorname{tr}(a^2+ub^2)=0
\right\}.
$  In \cite{9}, Wang et al. uses the general defining-set construction 
$
D=\{d_1,d_2,\ldots,d_n\}
\subseteq \operatorname{GR}(4,m)\setminus\{0\},
$
with the associated code
$
C_D=
\left\{
\big(\operatorname{Tr}(xd)\big)_{d\in D}
:\ x\in\operatorname{GR}(4,m)
\right\},
$ 
where $\operatorname{Tr}:\operatorname{GR}(4,m)\to\mathbb{Z}_4$
is the Galois-ring trace.\\\\
   The motivation of our work is mainly from the existing literatures on $Lee$ weight distributions, trace codes, few weight codes and mixed alphabet. We organizes and investigated different types of defining sets within a unified framework for the mixed alphabet $Z^{\alpha}_2 Z^{\beta}_4$ for finite value of $\alpha$ and $\beta$. Different choices of defining sets influences the $Lee$ weight distributions and the resulting properties. We organizes our work in 6 sections. In section 1, we establises our motivation of works over mixed alphabet. Section 2 gives some basics definitions that we have used. In Section 3, we studied A to J various kinds of defining sets for codes over the mixed alphabet with their useful results on Parseval identitity. In section 4, and section 5, we establises new kind of result on 2-$Lee$ weight distribution  and 3-$Lee$ weight distribution codes of $Z_2Z_4$ additive codes respectively. In section 6, we gave a conclusion.\\
   For convenience, we list the notations used in this paper as follows:\\
   
\begin{tabbing}
$\mathbb{C}^*$  \qquad \= \qquad the set of nonzero complex numbers, \\
$R$ \> \qquad the mixed ring $\mathbb{Z}_2^\alpha \mathbb{Z}_4^\beta$, where $\alpha$ and $\beta$ are finite integers, \\
$D$ \> \qquad any defining set from A to J, \\
$C_D$ \>\qquad the code corresponding to the defining set $D$, \\
$C_D(u)$ \> \qquad the codeword associated with the vector $u$, \\
$u \cdot t$ \>\qquad the inner product associated with the mixed-alphabet structure,\\
$W_L(u)$   \> \qquad    the $Lee$-weight of the vector $u$, \\
$d_L(C)$ \>\qquad the minimum $Lee$ distance of the code $C$, \\
$\Phi(u)$ \>\qquad the Gray image of the vector $u$, \\
$W_H(u)$ \> \qquad the Hamming weight of the vector $u$, \\
$H_D(u)$ \>\qquad  the character sum associated with the defining set $D$ and the vector $u$, \\
$H_{D^c}(u)$ \> \qquad the character sum associated with the complement of $D$, \\
$|u|$ \> \qquad the cardinality of the support of the vector $u$, \\
$N$ \> \qquad the length of the binary Gray image, \\
$\operatorname{Im}(T)$ \> \qquad the image of the linear operator $T$, \\
$W$ \> \qquad the cardinality of the defining set $D$,\\
$F(u)$\> \qquad  $W_L\big(C_D(u)\big)$.
\end{tabbing}
\section{preliminaries}
 A subset $C\subseteq Z^{\alpha}_2 Z^{\beta}_4$ is called a $Z_2Z_4$-additive code if $u+v\in C$ for $u,v\in C$. We will take the length of the code $C$ as $n=\alpha+2\beta$ in general. The additive code $C$ is called polycyclic, if there exists two vectors say $(c_0,c_1,...,c_{\alpha-1}) \in Z^{\alpha}_2$, $(c^{'}_0,c^{'}_1,...,c^{'}_{\beta -1})\in Z^{\beta}_4 $ and such that for any codeword $(v|w)=(v_0,v_1,...,v_{\alpha-1}|w_0,w_1,...,w_{\beta-1}) \in C$, its poly-shift $T(v|w)=\big((0,v_0,v_1...,v_{\alpha-2})+v_{\alpha-1}(c_0,c_1,...,c_{\alpha-1})|(0,w_0,w_1,...,w_{\beta-2})+w_{\beta-1}(c^{'}_0,c^{'}_1,...,c^{'}_{\beta -1})\big)\\ \in C$. For a vector $u=(x_1,x_2,...,x_\alpha|y_1,y_2,...,y_\beta)\in Z^{\alpha}_2 Z^{\beta}_4$, the standard Gray map from $Z^{\alpha}_2 Z^{\beta}_4$ to $Z^{\alpha+2\beta}_2$ is $\Phi(u)=\big(x_1,x_2,..,x_\alpha| \phi(y_1),\phi(y_2),...,\phi(y_\beta)\big)$, where $\phi$ is defined from $Z_4$ to $Z^2_2$ as $\phi(0)=(0,0), \phi(1)=(0,1),\phi(2)=(1,1),\phi(3)=(1,0)$. One can verify that the Gray map defined here forms an equivalent relation. And $Lee$-weight of a vector in $Z^{\alpha}_2 Z^{\beta}_4$ is defined as $W_L(u)=W_H\big(\Phi(u)\big)$, where $W_H$ is defined as Hamming weight of the vector in $Z^{\alpha+2\beta}_2$. Assume that $D$ be the defining set for the $Z_2Z_4$-additive code, then we define a additive code as  $C_D=\{C_D(u)=(u\cdot t)_{t\in D}\big| u \in Z^{\alpha}_2Z^{\beta}_4\}$, where $u\cdot t=2\sum_{i=1}^{\alpha} u_it_i \pmod{2}+\sum_{j=\alpha+1}^{\alpha+\beta}u_{\alpha+j}t_{\alpha+j} \pmod{4}$ for $u=(u_1,u_2,...,u_{\alpha}\big| u_{\alpha+1},...,u_{\alpha +\beta}), t=(t_1,t_2,...,t_{\alpha}\big| t_{\alpha+1},...,t_{\alpha+\beta})$. We will explore defining sets from A to J. Here, the $Lee$-weight of the codeword $c=(c_1,...,c_{\alpha}\big| c_{\alpha+1},...,c_{\alpha+\beta})\in C_D$ is defined as \[
\begin{aligned}
\operatorname{W}_{L}(c)
&=
\sum_{j=1}^{\alpha}
\frac{1-(-1)^{c_j}}{2}
+
\sum_{j=\alpha+1}^{\alpha+\beta}
\left(1-\operatorname{Re}\left(i^{c_j}\right)\right)\\
&=
\frac{\alpha}{2}+\beta
-\frac{1}{2}\sum_{j=1}^{\alpha}(-1)^{c_j}
-\operatorname{Re}\left(
\sum_{j=\alpha+1}^{\alpha+\beta}i^{c_j}
\right).
\end{aligned}
\]
The $Lee$-weight of the code \(C_D\), or equivalently its minimum
$Lee$ distance, is defined by
\[
d_L(C_D)
=
\min_{c\in C_D\setminus\{0\}}
\operatorname{wt}_L(c).
\]
Therefore,
\[
d_L(C_D)
=
\min_{c\in C_D\setminus\{0\}}
\left\{
\frac{\alpha}{2}+\beta
-\frac{1}{2}\sum_{j=1}^{\alpha}(-1)^{c_j}
-\operatorname{Re}\left(
\sum_{j=\alpha+1}^{\alpha+\beta}i^{c_j}
\right)
\right\}.
\]

Let $v=(v_1,v_2,...,v_n)\in Z^{\alpha}_2Z^{\beta}_4$, then support denoted by $supp(v)$ is defined by the set of non zero coordinate positions. The $Hamming$ weight $W_H(v)$ of $v$ is defined by the size of $supp(v)$.
\begin{lemma}
For every $x\in R$, the Gray map $\Phi$ preserves $Lee$-weight,
that is,
\[
W_H(\Phi(x))=W_L(x),
\]
where $W_H$ denotes the $Hamming$ weight on
$\mathbb{F}_2^{\alpha+2\beta}$ and $W_L$ denotes the
$Lee$-weight on $\mathbb{Z}_2^\alpha\mathbb{Z}_4^\beta$.
\end{lemma}

\begin{proof}
Let
\[
x=(a_1,\ldots,a_\alpha,b_1,\ldots,b_\beta)\in R.
\]
By the definition of $\Phi$,
\[
W_H(\Phi(x))
=\sum_{i=1}^{\alpha}W_H(a_i)
+\sum_{j=1}^{\beta}W_H(\phi(b_j)).
\]
Since $W_H(a_i)=W_L(a_i)$ for $a_i\in\mathbb{Z}_2$ and
$W_H(\phi(b_j))=W_L(b_j)$ for every $b_j\in\mathbb{Z}_4$,
we obtain
\[
W_H(\Phi(x))
=\sum_{i=1}^{\alpha}W_L(a_i)
+\sum_{j=1}^{\beta}W_L(b_j)
=W_L(x).
\]
Hence the result follows.
\end{proof}
\begin{definition}
Let $ R=\mathbb{Z}_2^\alpha\mathbb{Z}_4^\beta,$
 and $q$ is an odd positive integer. For each
$
t=(x_1,\ldots,x_\alpha,y_1,\ldots,y_\beta)
\in R,
$
the $q$-cyclotomic coset generated by $t$ is defined as
\[
C_t^{(q)}
=
\left\{q^j t:j=0,1,2,\ldots\right\},
\]
where
$
q^jt=
(q^jx_1,\ldots,q^jx_\alpha,
q^jy_1,\ldots,q^jy_\beta),
$
with the first $\alpha$ coordinates computed
modulo $2$ and the remaining $\beta$
coordinates computed modulo $4$.
\end{definition}
Let $R=\mathbb{Z}_2^2\times\mathbb{Z}_4^2$
and choose $q=3$ and consider the element $ t=(1,0,1,2)\in R,$
then, the $3$-cyclotomic coset generated by $t$ is
\[
C_t^{(3)}=\{t,3t,3^2t,\ldots\},
\]
where multiplication is performed coordinatewise,
modulo $2$ in the first two coordinates and modulo
$4$ in the last two coordinates. We have
$
3t=(3,0,3,6)
\equiv (1,0,3,2)
\pmod{(2,2,4,4)},
$
and
$
3^2t=3(1,0,3,2)
=(3,0,9,6)
\equiv (1,0,1,2)
\pmod{(2,2,4,4)}.
$
Therefore,
$
C_{(1,0,1,2)}^{(3)}
=
\{(1,0,1,2),(1,0,3,2)\}.
$
Hence, the cardinality of this cyclotomic coset is $2$.
\section{Various Defining Sets and Generating Functions for
\texorpdfstring{$R$}{R}}

 We investigate the following defining sets in detail and establish new results concerning their $Lee$-weight distributions:
 \begin{enumerate}
     \item[A.]$D_t=\{w\in Z^{\alpha}_2Z^{\beta}_4\big|\ W_L(w)=t\}$.
     \item[B.] $D_{t_1,{t_2}}=\{ w\in Z^{\alpha}_2Z^{\beta}_4| t_1 \leq W_L(w)\leq t_2\}.$
     \item[C.]  $D_U=\cup D_i$, where $D_i=\{w\in Z^{\alpha}_2Z^{\beta}_4| W_L(w)=i\}.$
     
     \item[D.] $D_t=\{w\in Z^{\alpha}_2Z^{\beta}_4 \big| |supp(w)|=t\}$, where $supp(w)=\{i| w_i \neq 0\}.$
     \item[E.] $D_{Tr}=\{ w\in Z^{\alpha}_2Z^{\beta}_4 | {Tr}^m(aw)=0\}$, where $Tr^{m}(w)= \sum_{j=0}^{m-1} (x^{2j}+2y^{2j}).$
     \item [F.] $D_Q=\{w\in Z_2^{\alpha}Z_4^{\beta}| Q(w)=0\}$, where $Q(w)$ is quadratic form.
     \item[G.] $
D_{\mathrm{cl}}
=
\{\xi_i\in\mathbb{Z}_2^\alpha\mathbb{Z}_4^\beta:
i\in T\},$ where $T$ is finite  union of $q$-cyclotomic cosets.
 
     \item [H.] $D_{inn}=\{w\in Z^{\alpha}_2Z^{\beta}_4| w\cdot u=0, \forall u \in Z^{\alpha}_2Z^{\beta}_4\}$, where $w\cdot u=2\sum_{i=1}^{\alpha} w_iu_i+ \sum_{j=1}^{\beta} w_{\alpha+j}u_{\alpha+j} \pmod{ 4}.$
     \item[I.] $D_v=v+H, H\subseteq Z^{\alpha}_2Z^{\beta}_4$, produces non-linear codes if $v\neq 0.$
     \item[J.] $D_s=\{w=(x,y)\in Z^{\alpha}_2Z^{\beta}_4| supp(x)\subseteq supp(y)\}, x\in Z^{\alpha}_2,y\in Z^{\beta}_4\}.$
 
 \end{enumerate}
 \textbf{  Note}: The above defining sets are subsets of $Z^{\alpha}_2Z^{\beta}_4.\\$
   Similar to the generating set for binary code \cite{5}, we can extend generating set for $Z^\alpha_2Z^\beta_4$ as follows:

Let us define map
\[
\chi_u:R\longrightarrow\mathbb{C}^{*},
\qquad
\chi_u(t)=i^{u\cdot t},
\]
which is an additive character of \(R\), where $R=Z^\alpha_2Z^\beta_4$. Equivalently,
\[
\chi_u(t)
=
\prod_{j=1}^{\alpha}(-1)^{u_jt_j}
\prod_{j=\alpha+1}^{\alpha+\beta}i^{u_jt_j}.
\]
For a subset \(D\subseteq R\), define the associated character sum by
\[
H_D(u)
=
\sum_{t\in D}\chi_u(t)
=
\sum_{t\in D}
\prod_{j=1}^{\alpha}(-1)^{u_jt_j}
\prod_{j=\alpha+1}^{\alpha+\beta}i^{u_jt_j}.
\]
We say that \(D\) is a generating set of \(R\) if
$
\langle D\rangle
=
\mathbb{Z}_2^\alpha\mathbb{Z}_4^\beta.
$
Since
$
D^c=R\setminus D,
$
we immediately have
$
H_D(u)+H_{D^c}(u)=H_R(u).
$
Now, since
$
R=\mathbb{Z}_2^\alpha\mathbb{Z}_4^\beta,
$
we obtain
$$
\begin{aligned}
H_R(u)
&=
\sum_{t\in R} i^{u\cdot t}\\
&=
\prod_{j=1}^{\alpha}
\left(
\sum_{t_j\in\mathbb{Z}_2}
(-1)^{u_jt_j}
\right)
\prod_{j=\alpha+1}^{\alpha+\beta}
\left(
\sum_{t_j\in\mathbb{Z}_4}
i^{u_jt_j}
\right).
\end{aligned}
$$
Therefore,
\[
H_D(u)+H_{D^c}(u)
=
\prod_{j=1}^{\alpha}
\left(1+(-1)^{u_j}\right)
\prod_{j=\alpha+1}^{\alpha+\beta}
\left(1+i^{u_j}+i^{2u_j}+i^{3u_j}\right).
\]
For \(u_j\in\mathbb{Z}_2\) for \(1\leq j\leq\alpha\) and
\(u_j\in\mathbb{Z}_4\) for \(\alpha+1\leq j\leq\alpha+\beta\), we have
\[
1+(-1)^{u_j}
=
\begin{cases}
2, & u_j=0,\\
0, & u_j=1,
\end{cases}
\]
and
\[
1+i^{u_j}+i^{2u_j}+i^{3u_j}
=
\begin{cases}
4, & u_j=0,\\
0, & u_j\neq0.
\end{cases}
\]
Consequently,
\[
H_R(u)
=
\begin{cases}
2^\alpha4^\beta, & u=0,\\
0, & u\neq0.
\end{cases}
\]
In particular, for every \(u\neq0\),
$
H_{D^c}(u)=-H_D(u).
$
\begin{lemma}[Parseval identity]
Let
$
R=\mathbb{Z}_2^\alpha\mathbb{Z}_4^\beta
$
and let \(D\subseteq R\). For \(u\in R\), define
$
H_D(u)=\sum_{t\in D}i^{u\cdot t},
$
where
$
u\cdot t
=
2\sum_{j=1}^{\alpha}u_jt_j\pmod{2}
+
\sum_{j=\alpha+1}^{\alpha+\beta}u_jt_j\pmod{4}.
$
Then
\[
\sum_{u\in R}|H_D(u)|^2
=
2^\alpha4^\beta |D|.
\]
\end{lemma}

\begin{proof}
We have
\[
\begin{aligned}
\sum_{u\in R}|H_D(u)|^2
&=
\sum_{u\in R}
H_D(u)\overline{H_D(u)}\\
&=
\sum_{u\in R}
\left(\sum_{t\in D}i^{u\cdot t}\right)
\left(\sum_{s\in D}i^{-u\cdot s}\right)\\
&=
\sum_{t,s\in D}
\sum_{u\in R}i^{u\cdot(t-s)}.
\end{aligned}
\]
By the orthogonality of the additive characters of \(R\), we have 
\[
\sum_{u\in R}i^{u\cdot(t-s)}
=
\begin{cases}
|R|, & t=s,\\
0, & t\ne s.
\end{cases}
\]
Since
$
|R|
=
|\mathbb{Z}_2^\alpha|
|\mathbb{Z}_4^\beta|
=
2^\alpha4^\beta,
$
it follows that
\[
\begin{aligned}
\sum_{u\in R}|H_D(u)|^2
&=
\sum_{t,s\in D}
\begin{cases}
2^\alpha4^\beta, & t=s,\\
0, & t\ne s
\end{cases}\\
&=
2^\alpha4^\beta |D|.
\end{aligned}
\]
Hence,
$
\sum_{u\in R}|H_D(u)|^2
=
2^\alpha4^\beta |D|.
$
\end{proof}

\begin{theorem}
Let
$
R=\mathbb{Z}_2^\alpha\mathbb{Z}_4^\beta,
$
and let $D\subseteq R$ be such that $D=-D$, where $D$ is defining set from A to J. Consider  
\[
C_D=\left\{C_D(u):u\in R\right\},
\]
where
$
C_D(u)=\left(u\cdot t\right)_{t\in D}
$
and
$
u\cdot t
=2\sum_{j=1}^{\alpha}u_jt_j\pmod{2}+ \sum_{j=\alpha+1}^{\alpha+\beta}u_jt_j \pmod{4}.
$
Then, for every $u\in R$,
\[
\operatorname W_L(C_D(u))
=
|D|-H_D(u).
\]
Moreover, if $A_w$ denotes the number of codewords of $Lee$-weight $w$ in $C_D$, then
\[
\sum_w A_w\bigl(|D|-w\bigr)^2
=
2^\alpha4^\beta|D|.
\]
\end{theorem}

\begin{proof}
For $x\in\mathbb{Z}_4$, the $Lee$-weight can be expressed as
\[
\operatorname W_L(x)
=
1-\operatorname{Re}(i^x).
\]
Indeed,
\[
\operatorname W_L(0)=0,\qquad
\operatorname W_L(1)=1,\qquad
\operatorname W_L(2)=2,\qquad
\operatorname W_L(3)=1,
\]
and the above identity holds for all $x\in\mathbb{Z}_4$.

For $u\in R$, we have
\[
C_D(u)=\left(u\cdot t\right)_{t\in D}.
\]
Therefore,
\begin{align*}
\operatorname W_L(C_D(u))
&=
\sum_{t\in D}
\operatorname W_L(u\cdot t)\\
&=
\sum_{t\in D}
\left(1-\operatorname{Re}
\left(i^{u\cdot t}\right)\right)\\
&=
|D|-\operatorname{Re}
\left(
\sum_{t\in D}i^{u\cdot t}
\right)\\
&=
|D|-\operatorname{Re}\left(H_D(u)\right).
\end{align*}

Since $D=-D$, for every $t\in D$, we also have $-t\in D$. Moreover,
\[
i^{u\cdot(-t)}
=
i^{-u\cdot t}
=
\overline{i^{u\cdot t}}.
\]
Thus, the terms in $H_D(u)$ occur in conjugate pairs, and consequently
\[
H_D(u)\in\mathbb{R}.
\]
Hence,
\[
\operatorname{Re}\left(H_D(u)\right)=H_D(u),
\]
and therefore
\[
\operatorname W_L(C_D(u))
=
|D|-H_D(u).
\]
Equivalently,
\[
H_D(u)
=
|D|-\operatorname W_L(C_D(u)).
\]

By Lemma~3.1 (Parseval identity),
\[
\sum_{u\in R}|H_D(u)|^2
=
2^\alpha4^\beta|D|.
\]
Since $H_D(u)$ is real, we obtain
\[
\sum_{u\in R}
\left(|D|-\operatorname{W}_L(C_D(u))\right)^2
=
2^\alpha4^\beta|D|.
\]

Finally, grouping the elements $u\in R$ according to the $Lee$-weight
\[
w=\operatorname W_L(C_D(u)),
\]
and denoting by $A_w$ the number of codewords of $Lee$-weight $w$, we obtain
\[
\sum_w A_w\left(|D|-w\right)^2
=
2^\alpha4^\beta|D|.
\]
This completes the proof.
\end{proof}
\begin{proposition}
Let $R=\mathbb{Z}_2^\alpha\mathbb{Z}_4^\beta$ and
\[
D_t=\{w\in R:|\operatorname{supp}(w)|=t\},
\]
where $\operatorname{supp}(w)$ denotes the set of nonzero
coordinates of $w$. Define
\[
H_{D_t}(u)=\sum_{w\in D_t}i^{w\cdot u},
\qquad u\in R,
\]
where $i=\sqrt{-1}$ and $w\cdot u$ is the standard
$\mathbb{Z}_2\mathbb{Z}_4$ inner product. Then
\[
\sum_{u\in R}|H_{D_t}(u)|^2
=
2^\alpha 4^\beta
\sum_{r=0}^{t}
\binom{\alpha}{r}
\binom{\beta}{t-r}3^{t-r}.
\]
\end{proposition}

\begin{proof}
By Parseval's identity for characters on $R$, we have
\[
\sum_{u\in R}|H_{D_t}(u)|^2
=|R|\,|D_t|.
\]
Since $|R|=2^\alpha 4^\beta$, it remains to determine
the cardinality of $D_t$.

Suppose that exactly $r$ of the $t$ nonzero coordinates
belong to the $\mathbb{Z}_2$ component. The remaining
$t-r$ nonzero coordinates belong to the $\mathbb{Z}_4$
component. We may choose these coordinates in
\[
\binom{\alpha}{r}\binom{\beta}{t-r}
\]
ways. Each nonzero coordinate in $\mathbb{Z}_2$ has
one possible value, whereas each nonzero coordinate in
$\mathbb{Z}_4$ has three possible values.

Consequently,
\[
|D_t|
=
\sum_{r=0}^{t}
\binom{\alpha}{r}
\binom{\beta}{t-r}3^{t-r}.
\]
Substituting this expression into Parseval's identity
yields
\[
\sum_{u\in R}|H_{D_t}(u)|^2
=
2^\alpha 4^\beta
\sum_{r=0}^{t}
\binom{\alpha}{r}
\binom{\beta}{t-r}3^{t-r}.
\]
This completes the proof.
\end{proof}
\begin{theorem}[Lee-weight distribution using defining sets A, B and C]
Let
\[
N=\alpha+2\beta.
\]
For $t\geq 0$, if we take $D$ as
$D_t$, $D_{t_1,t_2}$ and $D_U$ as defined in A., B. and C. respectively, then
\[
\operatorname W_L(C_D(u))
=
|D|-H_D(u),
\]
and, if $A_w^{D}$ denotes the number of codewords of \textit{Lee}-weight of $w$
in $C_D$, then
\[
\sum_w A_w^{D}(|D|-w)^2
=
2^\alpha4^\beta|D|.
\]
In particular,
\[
\sum_w A_w^{(t)}
\left(
\binom{N}{t}-w
\right)^2
=
2^\alpha4^\beta\binom{N}{t},
\]
\[
\sum_w A_w^{(t_1,t_2)}
\left(
\sum_{t=t_1}^{t_2}\binom{N}{t}-w
\right)^2
=
2^\alpha4^\beta
\sum_{t=t_1}^{t_2}\binom{N}{t},
\]
and
\[
\sum_w A_w^{(U)}
\left(
\sum_{i\in U}\binom{N}{i}-w
\right)^2
=
2^\alpha4^\beta
\sum_{i\in U}\binom{N}{i}.
\]
\end{theorem}

\begin{proof}
For an element
$
w=(w_1,\ldots,w_{\alpha+\beta})\in
\mathbb{Z}_2^\alpha\mathbb{Z}_4^\beta,
$
each binary coordinate contributes either $0$ or $1$ to the \textit{Lee}-weight. Each $\mathbb{Z}_4$ coordinate contributes $0$, $1$, or $2$,
with one element of weight $0$, two elements of weight $1$, and one
element of weight $2$. Hence the $Lee$-weight enumerator of $R$ is
\[
(1+x)^\alpha(1+2x+x^2)^\beta.
\]
Since
\[
1+2x+x^2=(1+x)^2,
\]
we obtain
\[
(1+x)^\alpha(1+2x+x^2)^\beta
=
(1+x)^{\alpha+2\beta}
=
(1+x)^N.
\]
Therefore, the coefficient of $x^t$ is
\[
|D_t|=\binom{N}{t}.
\]

Since the sets $D_t$ are pairwise disjoint,
\[
D_{t_1,t_2}
=
\bigcup_{t=t_1}^{t_2}D_t,
\]
and consequently
\[
|D_{t_1,t_2}|
=
\sum_{t=t_1}^{t_2}|D_t|
=
\sum_{t=t_1}^{t_2}\binom{N}{t}.
\]
Similarly,
\[
|D_U|
=
\sum_{i\in U}|D_i|
=
\sum_{i\in U}\binom{N}{i}.
\]

Next, observe that
\[
\operatorname W_L(-w)=\operatorname W_L(w)
\]
for every $w\in R$. Hence
\[
D_t=-D_t,
\qquad
D_{t_1,t_2}=-D_{t_1,t_2},
\qquad
D_U=-D_U.
\]
Thus, for any of these defining sets $D$, the terms corresponding to
$w$ and $-w$ in
$H_D(u)$ are complex conjugates. Hence $ H_D(u)\in\mathbb{R}.$

For $x\in\mathbb{Z}_4$, we have
\[
\operatorname W_L(x)
=
1-\operatorname{Re}(i^x).
\]
Therefore,
\begin{align*}
\operatorname W_L(c(u))
&=
\sum_{w\in D}
\operatorname W_L(u\cdot w)\\
&=
\sum_{w\in D}
\left(1-\operatorname{Re}(i^{u\cdot w})\right)\\
&=
|D|-\operatorname{Re}
\left(
\sum_{w\in D}i^{u\cdot w}
\right)\\
&=
|D|-H_D(u).
\end{align*}
Consequently,
\[
H_D(u)=|D|-\operatorname W_L(c(u)).
\]

By Lemma~3.1 (Parseval identity),
\[
\sum_{u\in R}|H_D(u)|^2
=
2^\alpha4^\beta|D|.
\]
Since $H_D(u)$ is real, it follows that
\[
\sum_{u\in R}
\left(
|D|-\operatorname{W}_L(c(u))
\right)^2
=
2^\alpha4^\beta|D|.
\]
Grouping the elements $u\in R$ according to the $Lee$ weight of
$C_D(u)$ gives
\[
\sum_w A_w^D(|D|-w)^2
=
2^\alpha4^\beta|D|.
\]

Substituting the explicit expressions for $|D_t|$,
$|D_{t_1,t_2}|$, and $|D_U|$ yields the three stated identities.
\end{proof}

\begin{ex}
Let $\alpha=1$ and $\beta=1$. Then
$
R=\mathbb{Z}_2\times\mathbb{Z}_4
\quad\text{and}\quad N=3.$
For $t=(x,y)\in R$, define the Lee weight by
$
\operatorname W_L(t)=x+w_L(y),
$
where $w_L$ denotes the $Lee$ weight in $\mathbb{Z}_4$.
The mixed inner product is given by
$
u\cdot t=2ax\pmod 2 +by\pmod 4,
$
where $u=(a,b)\in R$.

\textbf{Case 1.} Take
$
D_1=\{t\in R:\operatorname W_L(t)=1\}.
$
Then
$
D_1=\{(0,1),(0,3),(1,0)\}
$
and $|D_1|=\binom{3}{1}=3$.
The corresponding codeword is
$
C_{D_1}(u)=(b,3b,2a).
$
The $Lee$-weight distribution of $C_{D_1}$ is given by
\[
\begin{array}{|c|c|c|c|c|}
\hline
w&0&2&4&6\\ \hline
A_w^{(1)}&1&3&3&1\\
\hline
\end{array}
\]
Consequently,
\begin{align*}
\sum_w A_w^{(1)}(3-w)^2
&=1(3)^2+3(1)^2+3(-1)^2+1(-3)^2\\
&=24\\
&=2^1 4^1|D_1|.
\end{align*}

\textbf{Case 2.} Take
$
D_{1,2}=\{t\in R:1\leq
\operatorname W_L(t)\leq 2\}.\\
$
Then
\[
D_{1,2}=\{(0,1),(0,2),(0,3),(1,0),(1,1),(1,3)\}.
\]
Thus,
\[
|D_{1,2}|=\binom31+\binom32=6.
\]
The corresponding codeword is
\[
\begin{aligned}
C_{D_{1,2}}(u)=(&b,2b,3b,2a,
2a+b,2a+3b).
\end{aligned}
\]
The $Lee$-weight distribution is
\[
\begin{array}{|c|c|c|c|}
\hline
w&0&6&8\\ \hline
A_w^{(1,2)}&1&4&3\\
\hline

\end{array}
\]
Therefore,
\begin{align*}
\sum_w A_w^{(1,2)}(6-w)^2
&=1(6)^2+4(0)^2+3(-2)^2\\
&=48\\
&=2^1 4^1|D_{1,2}|.
\end{align*}

\textbf{Case 3.} Take $U=\{1,3\}$. Then
$
D_U=D_1\cup D_3,
$
where
$
D_3=\{(1,2)\}.
$
Hence,
\[
D_U=\{(0,1),(0,3),(1,0),(1,2)\}
\]
and
$$
|D_U|=\binom31+\binom33=4.
$$
The corresponding codeword is
$
C_{D_U}(u)=(b,3b,2a,2a+2b).
\\$
Their $Lee$-weight distribution is
\[
\begin{array}{|c|c|c|c|}
\hline
w&0&4&8\\ \hline
A_w^{(U)}&1&6&1\\
\hline
\end{array}
\]
Consequently,
\begin{align*}
\sum_w A_w^{(U)}(4-w)^2
&=1(4)^2+6(0)^2+1(-4)^2\\
&=32\\
&=2^1 4^1|D_U|.
\end{align*}
Thus, all three identities in Theorem~3.2
are verified for the above defining sets.
\end{ex}
\begin{table}[H]
\centering
\caption{Parseval identities for various defining sets over
$\mathbb{Z}_2^\alpha\mathbb{Z}_4^\beta$.}
\label{tab:parseval-defining-sets}
\renewcommand{\arraystretch}{1.5}
\begin{tabular}{|c|c|c|}
\hline
\textbf{Defining set} &
\textbf{Cardinality} &
\textbf{Parseval identity} \\ \hline

$D_t$ &
$\displaystyle
\sum_{r=0}^{t}
\binom{\alpha}{r}
\binom{\beta}{t-r}3^{t-r}
$ &
$\displaystyle
\sum_{u\in R}|H_{D_t}(u)|^2
=|R|\,|D_t|
$ \\ \hline

$D_{\mathrm{Tr}}$ &
$\displaystyle
\frac{|R|}{|\operatorname{Im}T_a|}
$ &
$\displaystyle
\sum_{u\in R}|H_{D_{\mathrm{Tr}}}(u)|^2
=\frac{|R|^2}{|\operatorname{Im}T_a|}
$ \\ \hline

$D_Q$ &
$N_Q$ &
$\displaystyle
\sum_{u\in R}|H_{D_Q}(u)|^2
=|R|N_Q
$ \\ \hline

$D_{\mathrm{cl}}$ &
$\displaystyle
\sum_{j=1}^{s}|C_j|
$ &
$\displaystyle
\sum_{u\in R}|H_{D_{\mathrm{cl}}}(u)|^2
=|R|\sum_{j=1}^{s}|C_j|
$ \\ \hline

$D_{\mathrm{inn}}$ &
$1$ &
$\displaystyle
\sum_{u\in R}|H_{D_{\mathrm{inn}}}(u)|^2
=|R|
$ \\ \hline

$D_v=v+H$ &
$|H|$ &
$\displaystyle
\sum_{u\in R}|H_{D_v}(u)|^2
=|R|\,|H|
$ \\ \hline

$D_s$ &
$\displaystyle
7^\alpha4^{\beta-\alpha}
$ &
$\displaystyle
\sum_{u\in R}|H_{D_s}(u)|^2
=|R|\,7^\alpha4^{\beta-\alpha}
$ \\ \hline

\end{tabular}
\end{table}
\begin{remark}
    Table 1 represent the Parseval identities of various kinds of defining sets.
\end{remark}
\begin{theorem}
Let $R=\mathbb{Z}_2^\alpha\mathbb{Z}_4^\beta$ and let
$D\subseteq R$ be any defining set from A to J in section 3. Define
\[
\begin{aligned}
\varphi_D:R&\longrightarrow\mathbb{Z}_4^{|D|},\\
u&\longmapsto c_D(u)=(u\cdot t)_{t\in D},
\end{aligned}
\]

Then
\[
|C_D|=\frac{2^{\alpha+2\beta}}{|\ker(\varphi_D)|}.
\]
\end{theorem}
\begin{proof}
By definition,
\[
C_D=\operatorname{Im}(\varphi_D).
\]

The mapping $\varphi_D$ is a group homomorphism. Indeed, for
any $u,v\in R$ and $t\in D$, we have
\[
\varphi_D(u+v)_t
=(u+v)\cdot t
\equiv u\cdot t+v\cdot t\pmod 4.
\]
Therefore,
\[
\varphi_D(u+v)=\varphi_D(u)+\varphi_D(v).
\]

The kernel of $\varphi_D$ is
\[
\begin{aligned}
\ker(\varphi_D)
&=\{u\in R:\varphi_D(u)=0\}\\
&=\{u\in R:u\cdot t=0
\text{ for all }t\in D\}\\
\end{aligned}
\]

Hence, by the First Isomorphism Theorem,
\[
R/\ker(\varphi_D)\cong C_D.
\]
Taking cardinalities, we obtain
\[
|C_D|=\frac{|R|}{|\ker(\varphi_D)|}.
\]
Since
\[
\begin{aligned}
|R|
&=|\mathbb{Z}_2^\alpha|\,
  |\mathbb{Z}_4^\beta|\\
&=2^\alpha 4^\beta\\
&=2^{\alpha+2\beta},
\end{aligned}
\]
we obtain
\[
|C_D|=\frac{2^{\alpha+2\beta}}{|\ker(\varphi_D)|}.
\]
This completes the proof.
\end{proof}
\section{2-Lee Weight Distribution of
\texorpdfstring{$\mathbb{Z}_2\mathbb{Z}_4$}
{Z2Z4} Additive Codes}
\begin{definition}
    For additive code $C_D$ over $Z^{\alpha}_2Z^{\beta}_4$ for a defining set $D$, the 2-$Lee$ weight code is defined as 
    \[
W_L(u)=
\begin{cases}
W,& u\neq0,\\
0,& u=0.     
\end{cases}
\]
    where $W$ is the cardinality of $D$.
\end{definition}
\begin{lemma}
    Let $D=\{t=(x,y)\in Z^{\alpha}_2Z^{\beta}_4 \big| supp(x)\subseteq supp(y), \big|supp(y)\big|=k,y_i\in \{0,2\}\}$, then the code $C_D$ is a 2-$Lee$ weight code. 
\end{lemma}
\begin{proof}
    Since $y_i\in \{0,2\}$, we can write as follows 
    \[
b_iy_i=
\begin{cases}
2b_i,& y_i=2,\\
0,& y_i=0.     
\end{cases}
\]  for all $i\in [1, \beta]$ where $b=(b_1,b_2,...,b_{\beta}),y=(y_1,y_2,...,y_{\beta})\in Z^\beta_4$. So, \[
b\cdot y \equiv 2\sum_{i\in\operatorname{supp}(y)} b_i \pmod{4}
\]
Therefore $b\cdot y\in \{0,2\}$. Now for $u=(a,b)\in Z^{\alpha}_2Z^{\beta}_4 $, we have  \[u\cdot t=2(a\cdot x)+ b\cdot y \pmod{4}=2\Big((x\cdot a)+ \sum_{i\in\operatorname{supp}(y)} b_i \pmod{4}\Big) .\]
Therefore  $u\cdot t \in \{0,2\}$.
As, $W_L(0)=0, W_L(2)=2$ in $Z_4,$\\ 
we have \[ W_L\big(C_D(u)\big)= 2 \times 
\left|
\left\{t\in D
: u\cdot t=2
\right\}
\right|.
\] Consider, \[\Phi_u(t)=(x\cdot a)+ \sum_{i\in\operatorname{supp}(y)} b_i \pmod{2},\] then \[u\cdot t= \begin{cases}
2,& \Phi_u(t)=1,\\
0,& \Phi_u(t)=0.   
\end{cases}\]
 Again for fixed $y$, there are $2^k$ possible values of $x$ and $a\cdot x$ takes values 0 and 1 equally. So,  $\Phi_u(t)=0$ and 1 each occurs $2^{k-1}$ times. Summing over all $y$, half of $D$ give value 0 and half give value 2. Hence\[W_L\big(C_D(u)\big)=\begin{cases}
0,& u=0,\\
|D|, & u\neq 0
\end{cases}
\]
 \end{proof}
 \begin{theorem}
Let
$
R=\mathbb{Z}_2^\alpha\mathbb{Z}_4^\beta,
$
and let $k$ be an integer satisfying $1\leq k\leq\beta$.
Define
\[
\begin{aligned}
D_k=\{(x,y)\in R:\;&
\operatorname{supp}(x)\subseteq\operatorname{supp}(y),\\
&|\operatorname{supp}(y)|=k,\quad
y_j\in\{0,2\},\ 1\leq j\leq\beta\}.
\end{aligned}
\]
Let $C_{D_k}$ be the code associated with $D_k$, defined as in the preliminary section. Then $C_{D_k}$ is a 2-$Lee$-weight code.
\end{theorem}

\begin{proof}
By the definition of $D_k$, every element $d=(x,y)\in D_k$
satisfies
\[
\operatorname{supp}(x)\subseteq\operatorname{supp}(y),
\qquad |\operatorname{supp}(y)|=k,
\]
and $y_j\in\{0,2\}$ for all $1\leq j\leq\beta$.
Therefore, $D_k$ satisfies the hypotheses of Lemma~4.1.
Applying Lemma~4.1, the associated code $C_{D_k}$ is a
2-$Lee$-weight code. Hence the theorem follows.
\end{proof}
 \begin{ex}
     Take $\alpha=2, \beta=3, k=2$ so that $D=\{(x,y)\big| supp(x)\subseteq supp(y), y_i\in \{0,2\}, \big|supp(y)\big|=2\}$. Choose $y\in Z^3_4$ so that exactly two positions are non zero. So possible supports are $\{1,2\}, \{1,3\}$ and $\{2,3\}$. So for each supports corresponding values of $y$ are $(2,2,0), (2,0,2)$ and $(0,2,2)$ respectively. As $supp(x) \subseteq supp(y), \big|supp(y)\big|=2$, we can choose $x\in Z^3_2$ and $x$ can have $1^{'s}$ only in those non zero positions. So, $x$ has 4 choices. Hence $\big|D\big|=4 \times 3=12$. In general, we can extend the cardinality of defining set of $D$ as $\binom{n}{k} 2^k$, if $\beta=n$ and $\big|supp(y)\big|=k$. For our particular choice of $\alpha, \beta $ and $k$, $W_L(C_D(u))\in \{0,12\}$. Since $u.t \in \{0,2\}$, for $v \in Z^\alpha_2Z^\beta_4$, the product $(u.t)(v.t) \in \{0,4\}\equiv 0 \pmod{4}$. Therefore in this case $C_D\subseteq C^\perp_D$.
 \end{ex} 
   \begin{proposition}
Let $D$ be a nonempty subset of
$R=\mathbb{Z}_2^\alpha\mathbb{Z}_4^\beta$, and let
$C_D$ be the code defined by
\[
C_D=\left\{c_a=(a\cdot d)_{d\in D}:a\in R\right\},
\]
where
\[
a\cdot d
=2\sum_{i=1}^{\alpha}a_i d_i
+\sum_{j=1}^{\beta}a_{\alpha+j}d_{\alpha+j}
\pmod 4.
\]
Suppose that the $Lee$-weight enumerator of $C_D$ is
\[
W_{C_D}(z)
=1+A_{w_1}z^{w_1}+A_{w_2}z^{w_2},
\]
where $0<w_1<w_2$ and $A_{w_1},A_{w_2}>0$.
Then $C_D$ is a 2-$Lee$-weight code with nonzero $Lee$
weights $w_1$ and $w_2$.
\end{proposition}

\begin{proof}
By definition, the $Lee$-weight enumerator of $C_D$ is
\[
W_{C_D}(z)=\sum_{c\in C_D}z^{W_L(c)},
\]
where $W_L(c)$ denotes the Lee weight of the codeword $c$.
The constant term $1$ corresponds to the zero codeword.
Since $A_{w_1}>0$ and $A_{w_2}>0$, there exist nonzero
codewords of $Lee$-weights $w_1$ and $w_2$, respectively.
Moreover, the assumed form of the weight enumerator shows
that no other nonzero $Lee$ weights occur in $C_D$.
Hence, $C_D$ has exactly two distinct nonzero $Lee$-weights,
namely $w_1$ and $w_2$. Therefore, $C_D$ is a
2-$Lee$-weight code.
\end{proof}  
 
\section{3-Lee Weight Distribution of
\texorpdfstring{$\mathbb{Z}_2\mathbb{Z}_4$}
{Z2Z4} Additive Codes}
For our convenient we redefine our weight function as \[F(u)=W_L\big(C_D(u)\big)=\sum_{t\in D} W_L(u\cdot t) .\] For each $k\in Z_4$, define $N_k(u)=\big| \{t \in D|u\cdot t=k\}\big|$. Then \[F(u)= \sum_{k\in Z_4} N_kW_L(k)\] We have $W_L(0)=0, W_L(1)=W_L(3)=1, W_L(2)=2$ in $Z_4$. Hence, $F(u)=N_1(u)+N_3(u)+2 N_2(u)$. We define a reduction map $\Psi:Z_4\rightarrow Z_3$ as $\Psi(x)\equiv x \pmod{3}$.
\begin{lemma}
    Let $R=Z^\alpha_2Z^\beta_4, a\in R\setminus \{0\}$ and define $D=\{t\in R\big| \Psi (a\cdot t ) \in S\}, S\subseteq \{0,1,2\}$. Then for $u\in R\setminus \{0\}$, the pair $\big( N_1(u)+N_3(u), N_2(u)\big)$ takes at most three distinct values.
\end{lemma}
\begin{proof}
    Define $A_{k,i}(u)=\Big| \{ t\in R\big| u\cdot t=k, \Psi(a.t)=i\}\Big|$. Then we can proof that \[N_k(u)=\sum_{i\in S} A_{k,i}(u) .\] Using decomposition of $D$ we can write $D=D_0\cup D_1 \cup D_2$, where $D_i=\{ t\in R| u\cdot t=k, \Psi(a.t)=i\} $. Hence, we can write \[N_k(u)=\{t\in D\big| u\cdot t =k\}=\cup_{i\in S} \{t\in D_i\big| u\cdot t=k\},\] where all $D_i, i=0,1,2$ are disjoint. Therefore \[ \Big| \{ t\in D\big| u\cdot t=k\}\Big|=\sum_{i\in S} \Big| \{ t\in D_i\big|u\cdot t=k \}\Big|.\]  Hence $\Big| \{t\in D_i|\big| u\cdot t=k\}\Big|=A_{k,i}(u)$. Therefore, we can conclude \[ N_k(u)=\sum_{i\in S} A_{k,i}(u).\] Thus \[F(u)=\sum_{k\in Z_4} W_L(k)\sum_{i\in S} A_{k,i}(u).\] Take $l_u(t)=u\cdot t\in Z_4, l_a(t)=a\cdot t \in Z_4$ and define $l:t\rightarrow \big(l_u(t),l_a(t)\big)$ which is a linear map from $R$ to $Z_4\times Z_4$. The values of $A_{k,i}(u)$ depends on the maps $l_u$ and $l_a$. We divide the space that contain $u\neq 0$ into three disjoint classes as follows:
    \begin{itemize}
        \item Class I. $U_1=\{u\neq 0\big| u\cdot a=0\}$
        \item Class II. $U_2=\{u=\lambda a \big| \lambda \in Z^*_4\}$
        \item Class III. $U_3=R\setminus \big( U_1\cup U_2 \cup \{0\}\big).$
    \end{itemize}
    
    We shall prove that each class produces constant values of the pair $\big( N_1(u)+N_3(u), N_2(u)\big)$.
    \begin{itemize}
        \item Class I. Let $u\in U_1, i.e., u\cdot a=0$. For any $t\in R,$ consider $t^{\prime}=t+\lambda a, \lambda \in Z_4$. We have $l_u(t^{\prime})=u\cdot t^{\prime} +\lambda(u\cdot a)=u\cdot a=l_u(t)$ as $u\cdot a=0$. Hence $l_u(t)$ is constant on each coset $t+\langle a\rangle.$ Again $l_a(t^{\prime})=a\cdot t^{\prime}=a\cdot t+\lambda(a\cdot a)$. Now $\Psi(a\cdot t^{\prime})=\Psi(a\cdot t)+ \lambda c \pmod{3}$, where $c=\Psi(a\cdot a)\in Z_3, \lambda\in Z_4$. If $c=0$, then $\Psi(a\cdot t^{\prime})=\Psi(a\cdot t)$. So varying $\lambda$ does not move between different $D_i$. Hence whole orbit $t+\langle a\rangle$ stays inside one of $D_i$. If $c=1 \in Z^*_3$, then $\psi(a\cdot t^{\prime})$ takes values $r, r+1,r+2$ for $\lambda=0,1,2$ respectively, where $r=\Psi(a\cdot t)$. Similarly for $c=2$, $\Psi(a.t^{\prime})$ takes the values $r, r+2, r+1$ for $\lambda=0,1,2$ respectively. Hence the orbit $t+\langle a\rangle$ intersects all three sets $D_0,D_1,D_2$. Thus $A_{k,i}(u)$ is constant. So, $N_k(u)=\big|S\big| A_{k,i}(u)$. Hence $\big( N_1(u)+N_3(u), N_2(u)\big)= \big(A_1,A_2\big)$, where $A_1$ and $A_2$ are two independent constant. 
        \item Class II. If $u\in U_2, i.e.,u=\lambda a$ and $u\cdot t=\lambda(a\cdot t)$. If $t\in D_i$, then $\Psi(a.t)=i$. Thus, $u\cdot t=\lambda(a\cdot t)$ which depends only on $i$. Hence \[A_{k,i}(u)=\begin{cases}
|D_i|,& k=\lambda(a\cdot t),\\
0, otherwsise
\end{cases}.\] 
Thus \[N_k(u)= \sum_{i\in S} A_{k,i}\] depends only on sizes of $D_i$. So we can conclude in this case $\big(N_1(u)+N_3(u), N_2(u)\big)=\big(B_1,B_2\big)$, where $B_1$ and $B_2$ are two independent constants.
\item Class III. Consider $u\in U_3, i.e., u\cdot a\neq0, u \notin \langle a\rangle$. Here $l_u$ and $l_a$ are linearly independent maps. Define $\Phi(t)=\big( u\cdot t, a\cdot t\big) \in Z_4 \times Z_4$. Since $u$ and $a$ are independent, $\Phi$ has maximal element of rank 2. Hence, we can conclude $\big( N_1(u)+N_3(u), N_2(u)\big)=\big(C_1,C_2\big)$, where $C_1$ and $C_2$ are two independent constants.
    \end{itemize}
    \end{proof}
    \begin{theorem}
        Let $R=Z^\alpha_2Z^\beta_4, a\in R\setminus \{0\}$, $a$ is fixed. Define the reduction mapping $\Psi:Z_4 \rightarrow Z_3 $, $\Psi(x)\equiv x \pmod{3}$. Define $D_i=\{t\in R\big| \Psi(a\cdot t)=i, i=0,1,2\}$. Let $C_D=\{C_D\big(u\big)=(u\cdot t)_{t\in D}\big| u \in R\}$, where $D\in \{D_1,D_1\cup D_2, D_0\cup D_2\}$. Then $\Big|\{W_L\big(C_D\big)(u)\big| u\neq 0\}\Big| \leq 3$.
    \end{theorem}
   \begin{proof}
       The partition $R=D_0\cup D_1\cup D_2$ is well defined: Consider $\Phi(t)=\Psi(a\cdot t)$, which is well defined in $Z_3$. Define $D_i=\Phi^{-1}(i)=\{t\in R\big| \Phi(t)=i\}$. Since $\Phi(t)\in \{0,1,2\}$, there exists a unique $i\in \{0,1,2\}$ such that $\Phi(t)=i$. Hence by difinition of preimage, $t\in \Phi^{-1}(i)=D_i$. Thus $R\subseteq D_0\cup D_1 \cup D_2$. Hence $R=D_0\cup D_1 \cup D_2$ as $D_i\subseteq R$ for all $i$. Clearly all $D^{'s}_i$ are disjoint because of equivalence relation of $\Phi$. Define the weight function as \[F(u)=W_L\big( C_D(u)\big)=\sum_{t \in D} W_L(u\cdot t).\] For each $k \in Z_4$, define $N_k=\Big|\{ t \in D\big| u\cdot t=k\}\Big|$. Then \[ F(u)=\sum_{k\in Z_4} N_k(u)W_L(u).\] As $W_L(0)=0,W_L(1)=W_L(3)=1, W_L(2)=2$ in $Z_4$, hence we have $F(u)=N_1(u)+N_3(u)+2N_2(u)$. Now using $Lemma$ 5.1, we can conlude $F(u)$ is completely determine by the triple $N_1(u)+N_3(u), N_2(u)$. Hence  $\Big|\{W_L\big(C_D\big)(u)\big| u\neq 0\}\Big| \leq 3$.
   \end{proof}

   \begin{ex}
For $u=(r,s)\in R$, the codeword corresponding to $u$ is
\[
C_D^u=(u\cdot t)_{t\in D},
\]
where inner product is defined as $ u\cdot(x,y)=2rx+sy\pmod 4.$ Let fix
$a=(1,1)\in R\setminus\{0\}$. For $t=(x,y)\in R$, the
mixed product is defined by $ a\cdot t=2x+y\pmod 4.$
Hence,
$
\Psi(a\cdot t)=\Psi(2x+y),
$
where $\Psi:\mathbb{Z}_4\to\mathbb{Z}_3$ is the reduction
mapping given by $\Psi(z)\equiv z\pmod 3$. Evaluating $\Psi(a\cdot t)$ for every $t\in R$, we obtain
the following table:
\[
\begin{array}{|c|c|c|}
\hline
t=(x,y) & a\cdot t\pmod 4 & \Psi(a\cdot t)\\
\hline
(0,0)&0&0\\
(0,1)&1&1\\
(0,2)&2&2\\
(0,3)&3&0\\
(1,0)&2&2\\
(1,1)&3&0\\
(1,2)&0&0\\
(1,3)&1&1\\
\hline
\end{array}
\]

Consequently, the defining sets are
\begin{align*}
D_0&=\{(0,0),(0,3),(1,1),(1,2)\},\\
D_1&=\{(0,1),(1,3)\},\\
D_2&=\{(0,2),(1,0)\}.
\end{align*}
Clearly, $R=D_0\cup D_1\cup D_2$, where the three sets
are pairwise disjoint.

We now construct the codes for the three choices of $D$
specified in $Theorem$~5.1.

\medskip
\noindent\textbf{Case 1: $D=D_1$.}
Ordering the elements of $D_1$ as $(0,1),(1,3)$, we obtain
\[
C_{D_1}^{(r,s)}=(s,2r+3s)\pmod 4.
\]
Therefore, $
C_{D_1}
=\left\{(s,2r+3s):r\in\mathbb{Z}_2,\
s\in\mathbb{Z}_4\right\}.
$
The Lee weights of the nonzero codewords are given by
\[
\begin{array}{|c|c|c|}
\hline
u & C_{D_1}^{u} & W_L(C_{D_1}^{u})\\
\hline
(0,1)&(1,3)&2\\
(0,2)&(2,2)&4\\
(0,3)&(3,1)&2\\
(1,0)&(0,2)&2\\
(1,1)&(1,1)&2\\
(1,2)&(2,0)&2\\
(1,3)&(3,3)&2\\
\hline
\end{array}
\]
Thus,
\[
\{W_L(C_{D_1}^{u}):u\in R\setminus\{0\}\}
=\{2,4\},
\]
and consequently,
\[
\left|
\{W_L(C_{D_1}^{u}):u\ne0\}
\right|=2\leq 3.
\]

\medskip
\noindent\textbf{Case 2: $D=D_1\cup D_2$.}
Ordering the elements as $(0,1),(1,3),(0,2),(1,0)$, we obtain
\[
C_{D_1\cup D_2}^{(r,s)}
=(s,2r+3s,2s,2r)\pmod 4.
\]
Hence,
\[
C_{D_1\cup D_2}
=\left\{
(s,2r+3s,2s,2r):
r\in\mathbb{Z}_2,\ s\in\mathbb{Z}_4
\right\}.
\]

\medskip
\noindent\textbf{Case 3: $D=D_0\cup D_2$.}
Ordering the elements as
$(0,0),(0,3),(1,1),(1,2),(0,2),(1,0)$, we obtain
\begin{align*}
C_{D_0\cup D_2}^{(r,s)}
={}&(0,3s,2r+s, 2r+2s,2s,2r)\pmod 4.
\end{align*}
Consequently,
\begin{align*}
C_{D_0\cup D_2}
=\big\{&(0,3s,2r+s,2r+2s,2s,2r):\\
& r\in\mathbb{Z}_2,\ s\in\mathbb{Z}_4\big\}.
\end{align*}

Thus, the defining sets and corresponding codes have been
constructed explicitly for all three choices of $D$ in
$Theorem$~5.1. In particular, the $Lee$-weight calculation for
$D=D_1$ verifies the asserted bound in this case.
\end{ex}
 
\begin{table}[htbp]
\centering
\caption{Defining sets and $Lee$-weight distributions for
$R=\mathbb{Z}_2\times\mathbb{Z}_4$, where $a=(1,1)$.}
\label{tab:example51}
\renewcommand{\arraystretch}{1.3}
\begin{tabular}{|c|c|c|c|}
\hline
Defining set & Elements of $D$ & Lee weights
& No. of distinct weights \\
\hline
$D_1$
&
$\{(0,1),(1,3)\}$
&
$\{2,4\}$
&
$2$
\\
\hline
$D_1\cup D_2$
&
$\{(0,1),(1,3),(0,2),(1,0)\}$
&
$\{4,6\}$
&
$2$
\\
\hline
$D_0\cup D_2$
&
$\{(0,0),(0,3),(1,1),(1,2),(0,2),(1,0)\}$
&
$\{4,6\}$
&
$2$
\\
\hline
\end{tabular}
\end{table}
\begin{definition}
    Let $S_{\alpha}$ is symmetric group of $\alpha$ elements from $Z_2$ and $S_{\beta}$ be symmetric group of $\beta $ elements from $Z_4$. Define group under composition operation to be the collection of all maps of the form $g\big(x_1,x_2,...,x_\alpha\big| y_1,y_2,...,y_\beta\big)=\big( x_{\sigma(1)},...x_{\sigma(\alpha)}\big| \epsilon_1y_{\tau(1)},...,\epsilon_\beta y_{\tau(\beta)}\big)$ on $R=Z^\alpha_2Z^\beta_4$, where $\sigma\in S_\alpha, \tau\in S_\beta$ and $\epsilon_j\in \{1,-1\} \subseteq Z_4, \forall j\in [1,\beta]$.
\end{definition}
\begin{theorem}

Let $R=Z^\alpha_2Z^\beta_4$ and $\{1,2,...,\alpha+\beta\}=S_1\cup S_2$ a disjoint partition. Take $D_0=\{0\}$, $D_1=\{t\in R\big| supp(t)\in S_1\}$ and $D_2=\{ t\in R\big| supp(t)\in S_2\}$. For $D\in \{D_1, D_1\cup D_2\}$ and code $C_D=\{ C_D(u)=(u\cdot t)_{t\in D} \big| u \in R\}$, the $Lee$ function depends only on $\big| supp(u)\big|$. Moreover, if the partition $S_1, S_2$ induces exactly three distinct interaction numbers, then $C_D$ is a three $Lee$ weight code. 
\end{theorem}
\begin{proof}
    We have $u\cdot t= 2 (u_1\cdot x) + u_2\cdot y$ for $u=(u_1,u_2), t=(x,y) \in R$. Consider, $supp(u) \cap supp(t) = \phi$ which gives $u\cdot t=0.$ Thus $u\cdot t $ depends only on co-ordinates belonging to $supp(u) \cap supp(t).$ Hence consequently, $W_L(u\cdot t)$ also depends only on the intersection pattern between $supp(u)$ and $supp(t)$ which is invariant under purmutation. Let $G$ be the group generated by transformations, then $g(D_i)=D_i, \forall g\in G$. Consider $u,v\in R$ satisfying $\big|supp(u)\big|=\big|supp(v)\big|$, then we can have a one to one corresponding $g\in G$ such that $g(u)=v.$\\
     Now \[F(v)= \sum_{t\in D} W_L(v \cdot t)=\sum_{t\in D} W_L\big(g(u)\cdot t\big) \] 
     Now $g(u) \cdot t=u\cdot g^{-1}(t)$(using invariance of the inner product under $G$). Therefore \[F(v)=\sum_{t\in D} W_L \big(u\cdot g^{-1}(t) \big) .\] Since $g(D)=D$, changing variable $s=g^{-1}(t)$ gives \[F(v)= \sum_{s\in D} W_L\big( u\cdot s\big)=F(u).\] Therefore $F(u)$ depends only on $\big| supp(u)\big|.$ \\
     For fixed $u$, consider $I(u,t)=\big| supp(u) \cap supp(t)\big|$ and $F(u)$ depends on the possible values of $I(u,t)$. Here, the partition $S_1\cup S_2$ which induces exactly there possible intersection values namely $A\cap S_1=A$, or $A\cap S_2=A$ or $A=(A\cap S_1) \cup (A\cap S_2)$ respectively for $A \in R$. Thus $I(u,t)\in \{A\cap S_1, A\cap S_2,A\cap S_1) \cup (A\cap S_2) \}$. Hence we can conclude $F(u)$ takes exactly three distinct values as $u\neq 0$. Hence $C_D$ is a three $Lee$- weight code.
     
\end{proof}
\begin{ex}
    Let $R=Z^2_2  Z^2_4$, and consider the partition $S_1=\{1,2\}$ and $S_2=\{3,4\}$ of the positional set $\{1,2,3,4\}$. Let $D=D_1\cup D_2$ , where $D_i=\{t\in R\big| supp(t)\in S_i\}.$ Then we have $\big|D_1\big|=3$ and $\big|D_2\big|=15$ and as $D_i^{'s}$ are disjoint, so $\big|D\big|=18.$ For $u\neq 0\in R$, use $C_D(u)=(u.t)_{t\in D},$ where $u.t=2(u_1t_1+u_2t_2)+u_3t_3 \pmod{4}$ for $t=(t_1,t_2,t_3,t_4)\in R.$ The $Lee$ weight of $C_D(u)$ depends only on the support type $\big( \big|supp(u)\cap S_1\big|,\big|supp(u)\cap S_2\big|).$ The resulting \textit{Lee}-weight distribution is as follows:\\
    \begin{table}[ht]
\centering
\caption{$Lee$-weight distribution of $C_D$}
\label{tab:lee-weight-distribution}
\begin{tabular}{|c |c| c| c| c|}
\hline
$w_L$ & $0$ & $4$ & $16$ & $20$ \\ 
\hline
$A_{w_L}$ & $1$ & $3$ & $15$ & $45$ \\ 
\hline
\end{tabular}
\end{table}
 \[
W_{C_D}(z)
=1+3z^4+15z^{16}+45z^{20}
\] and $C_D$ is a three $Lee$-weight code.
\end{ex}

\section{Conclusion remarks }

In this paper, we investigate various kinds of defining sets, labelled A to J, for codes over the mixed-alphabet space
$\mathbb{Z}_2^\alpha \mathbb{Z}_4^\beta$, with particular emphasis on their $Lee$-weight distributions and associated Parseval identities. We establish a general formula for the $Lee$-weight distribution associated with each of the considered defining sets. Furthermore, the corresponding Parseval identities have been derived, providing a unified framework for studying the weight distributions of these codes. In particular, we establish new theorems concerning the construction and characterization of codes with 2-$Lee$-weight and 3-$Lee$-weight distributions. These results extend the existing approaches to the study of defining-set constructions and provide useful algebraic tools for investigating the structural properties of codes over mixed alphabets.

There are several interesting directions for future research. One promising direction is to investigate the construction of optimal and quantum error-correcting codes by exploiting the defining sets and weight-distribution results developed in this paper, together with other suitable algebraic techniques. Moreover, the present study is restricted to the mixed-alphabet space $\mathbb{Z}_2^\alpha \mathbb{Z}_4^\beta$. It would therefore be worthwhile to extend the proposed framework to mixed-alphabet structures involving more than two alphabets and to explore suitable generalizations of the defining sets considered in this work. Such investigations may lead to new classes of few-weight codes and potentially contribute to the construction of optimal binary and quantum codes. The above defining sets also can be use to extend the work presented in \cite{11}.\\\\


\begin{thebibliography}{0}

\bibitem{1} J. Borges, J. Rifà, J. M. Pujol, The structure of $Z_2Z_4$ additive codes. IEEE Transactions on Information Theory, \textbf{53(10)}, 3877–3891, 2007.
\bibitem{2} L. Zongbing, K. Cheng, The \textit{Lee}-weight distributions of a class of quaternary codes. IEEE Access, \textbf{11}, 60950-60956, 2023.
\bibitem{3} Y. Wu, C. Li, L. Zhang and F. Xiao, Quaternary Codes and Their Binary Images. IEEE Transactions on Information Theory, \textbf{70(7)}, 4759-4768, 2024.
\bibitem{4} M. Bilal, J. Borges, S. T. Dougherty, C. F. Córdoba, Optimal codes over $Z_2\times Z_4$. Jornadas de Matemática Discreta y Algorítmica, 131-139, 2010.
\bibitem{5}
J. Y. Hyun, H. K. Kim, Y. Wu and Q. Yue,
Optimal minimal linear codes from posets,
\textit{Designs, Codes and Cryptography},
\textbf{88}, 2475-2492, 2020.
\bibitem{6} J. Borges, S. T. Dougherty, C. Fernández-Córdoba, Characterization and constructions of self-dual codes over \(\mathbb Z_2\times\mathbb Z_4\). Advances in Mathematics of Communications, \textbf{ 6(3)}, 287–303, 2012.
\bibitem{7} J. Geng, H. Wu, P. Solé, On one \textit{Lee}-weight and two \textit{Lee }weight \(\mathbb Z_2\mathbb Z_4[u]\) additive codes and their constructions. Advances in Mathematics of Communications, \textbf{17(5)}, 1196–1209, 2023.
\bibitem{8} P. Kumar, N. M. Khan, Five \textit{Lee}-weight linear codes over \(\mathbb F_q+u\mathbb F_q\). Advances in Mathematics of Communications, \textbf{19(5)}, 1417–1432, 2025.
\bibitem{9} Z. Wang, N. Li, X. Zeng, X. Tang, Trace codes over \(\mathbb Z_4\) and their \textit{Lee}-weight distributions. IEEE Transactions on Information Theory, \textbf{72(2)}, 1051–1066, 2026.
\bibitem{10} Y. Liu, M. Shi, P. Solé, Two weight and three weight codes from trace codes over \(\mathbb F_p+u\mathbb F_p+v\mathbb F_p+uv\mathbb F_p\). Discrete Mathematics, \textbf{341(2)}, 350–357, 2018.
\bibitem{11} S. Das, M. K. Patel and B. Singh,
Quantum codes from $(\alpha_1+u\alpha_2+u\alpha_3+u\alpha_4)$-constacyclic
codes over the non-chain ring $F_q+uF_q+vF_q+uvF_q$ with $q$ is a
square of prime $p$.
Contemporary Mathematics, CONM17120 (Preprint).
\end{thebibliography}
\end{document}